\documentclass[%
 reprint,
 amsmath,amssymb,
 aps,
]{revtex4-2}

\usepackage{graphicx}
\usepackage{dcolumn}
\usepackage{bm}
\usepackage{hyperref}
\usepackage[usenames,dvipsnames]{color}
\usepackage{orcidlink}

\begin{document}

\preprint{APS/123-QED}

\title{Testing the hypothesis of a matter density discrepancy within $\Lambda$CDM model using multiple probes.}

\newcommand{\orcid}[1]{\orcidlink{#1}}

\author{Ziad Sakr \orcid{0000-0002-4823-3757}}
 \email{ziad.sakr@net.usj.edu.lb}
\affiliation{Institut f\"ur Theoretische Physik, University of Heidelberg, Philosophenweg 16, 69120 Heidelberg, Germany
}%
\affiliation{IRAP, Université de Toulouse, CNRS, CNES, UPS, Toulouse, France}
\affiliation{%
 Universit\'e St Joseph, Faculty of Sciences, Beirut, Lebanon
}%


\date{\today}

\begin{abstract}
We investigate whether the two cosmological discrepancies on the Hubble constant ($H_0$) and the matter fluctuation parameter ($\sigma_8$)  are suggesting and compatible with the existence of an additional one on the present value of the matter density ($\Omega_{\rm{M}}$). Knowing that the latter effects on observables is degenerate with those coming from $H_0$ and $\sigma_8$, we combined different probes in a way to break these degeneracies while adopting the agnostic approach of, either relaxing the calibration parameters in each probe in order to be set by the data, or by only including priors with the condition that they are obtained independently from the discrepant parameters. We also compiled and used a dataset from previous direct measurements of $\Omega_{\rm{M}}$ obtained in a model independent way using the Oort technique. We found when combining, as our baseline, galaxy cluster counts + cluster gas mass fraction probe + cosmic chronometers + direct $\Omega_{\rm{M}}$ + priors from BBN and CMB, that both parameters, $H_0$ and $\sigma_8$, are consistent with those inferred from local probes, with $\sigma_8 = 0.745 \pm 0.05$  while $H_0 = 73.8 \pm 3.01$, and that for a value of $\Omega_{\rm{M}} = 0.22 \pm 0.01$ at more than 3$\sigma$ from that usually determined by the cosmic microwave background (CMB). We also found similar preferences when replacing cosmic chronometers (CC) by the supernovae (SN) data while allowing its calibration parameter to vary. However discrepancies appeared when we combined SN in addition to CC suggesting either inconsistencies between the SN sample and the other probes used or a serious challenge to our hypothesis.
To further investigate the later, we performed some stress tests by adding constraints from the baryonic acoustic oscillations (BAO) and found that $H_0$ reverts back to lower values at the expense however of a value of $\sigma_8$ non compatible with its local inferred ones, while only a prior on the matter density obtained from the CMB data keeps $\sigma_8$ within the values usually obtained when adopting the calibration parameters of the low redshift growth of structures probes. We conclude from our adopted analysis that, either both tensions are compatible with the local inferred low values of matter density at odd with those obtained by CMB, reviving by then an overlooked discrepancy, or that further evidences are indicating that the $\Lambda$CDM model is facing 
more difficulties to accommodate simultaneously all the current available observations.
\end{abstract}

\maketitle


\section{Introduction and motivations}

Since its early establishment, the standard cosmological model $\Lambda$CDM was essentially one in which the 'dark energy' component was needed to account for a big discrepancy between, early background expansion observations showing consistency with the Einstein de Sitter model where matter density is equal to unity $\Omega_{\rm{M}} \sim 1.0$, and structure formation observations \cite{Ostriker:1995su,Dodelson:1999am} as well as a variety of other measurements of the matter density in the 1980s and 1990s \citep{Peebles:1984ge,Efstathiou:1990xe,Krauss:1995yb} inferring much lower values close to  $\Omega_{\rm{M}} \sim 0.3$. The discovery of the acceleration of the universe \cite{SupernovaSearchTeam:1998fmf,SupernovaCosmologyProject:1998vns} confirmed the cosmological constant as the best solution to this discrepancy and turns out to fit almost all the other subsequent observations even when measurements improved by one to two order of magnitudes from the time of the acceleration discovery. However, nowadays, with the proliferation of probes and the further increase in precision, the cosmological model is facing again, with more or less strong statistical evidence, several tensions between its parameters (see \cite{Perivolaropoulos:2021jda} for a review or \cite{Kroupa:2012qj} for issues on structures formation at small scales in $\Lambda$CDM). Here we focus on the two most commonly pertinent ones, the $H_0$ tension which is the $\sim\,4.2\sigma$ discrepancy between the local measurements of the Hubble constant \citep{2022ApJ...934L...7R} and its inferred value from the cosmic microwave background (CMB) spectrum data within the $\Lambda$CDM framework, and the milder $\sigma_8$ tension, where its local measurements, notably from cluster counts and weak lensing correlations is implying less clumpiness in matter distribution compared to the CMB spectrum derived value \citep{2014A&A...571A..20P,2016A&A...594A..24P,DES:2021wwk,KiDS:2020ghu} but also other probes such as \cite{Esposito:2022plo,Nunes:2021ipq}. 
Many theoretically based solutions were proposed and investigated to alleviate these tensions \citep[see][for a comprehensive review]{DiValentino:2021izs,Abdalla:2022yfr} without a truly convincing positive outcome.  Sakr et al. (2021) in a series of articles \cite{Sakr:2018new,Ilic:2019pwq,Sakr:2021jya} showed
that the three most common extensions to $\Lambda$CDM, i.e. a change in the equation of state of dark
energy or a change in the growth index parameter or adding massive neutrinos, were unable
of solving the $\sigma_8$ discrepancy, in particular when data from the baryonic acoustic oscillations (BAO) signature on galaxy clustering is included. The latter strongly ties the sound horizon of the CMB at early redshift to its own at late times \cite{Jedamzik:2020zmd} ruling out as well late time modifications to $\Lambda$CDM as solutions to the $H_0$ tension. Though early time solutions have been found to strongly reduce the $H_0$ tension to an acceptable one or two $\sigma$ level, it turns out that they exasperate the $\sigma_8$ tension as shown in \cite{Hill:2020osr} for example or more exhaustively in \citep{Schoneberg:2021qvd} where an assessment of different early time solutions was conducted without reaching an absolute winner.
Here we propose to investigate whether the inability of the different attempts to alleviate both tensions at once is suggesting the existence of an overlooked tension on the value of the matter density of the universe that needs to be fixed as well. Already, some insightful studies such as \cite{Wagner:2022etu} noted that a change in $\Omega_{\rm{M}}$ could have an impact on both discrepancies, or the work of \cite{Heisenberg:2022gqk} which assessed
the theoretical implications needed to solve both tensions and found a shift required in the value
of $\Omega_{\rm{M}}$. More recently, \cite{Smith:2022iax} constructed scaling relations between the cosmological parameters
inferred from CMB and showed that a low prior on $\Omega_{\rm{M}}$ indicates that if the latter decreases then the inferred value of $h$ will increase while \cite{2022arXiv220505017B} showed that the matter density inferred with local Hubble constant priors is at odd with that when using the $H_0$ inferred value from CMB data. This was also suggested more recently in \cite{Colgain:2022nlb,Dainotti:2022bzg} but also earlier in \cite{Sakr:2021jya} study mentioned above which noted in its conclusion that one of the reasons that the three $\Lambda$CDM extensions
are unable of fixing the tension was that they require values for $\Omega_{\rm{M}}$ and $H_0$ far from those in agreement
with present datasets.
In this work we want to complement and follow on this hypothesis, by combining different probes in a way to break these degeneracies while adopting the agnostic approach of, either relaxing the calibration parameters in each probe in order to be set by the data, or by only including priors with the condition that they are obtained independently from the discrepant parameters. To cement them all, we shall combine and use a dataset of compiled previous direct measurements of $\Omega_{\rm{M}}$ obtained in a model independent way using the Oort technique \cite{Oort:1958nna} in which light to mass ratio in clusters of galaxies with respect to that obtained from the background galaxies is considered as a direct proxy to the matter density since it translates a feature in the formation of structures that is only function of the matter content and not dependent
on the other cosmological parameters. This could provide a way, as our aim is here, to put constraints on the matter density, $H_0$ and $\sigma_8$ outside the grip of the CMB constraints.\\
This paper is organised as follows: in Sect.~\ref{sec:datamod} we present the pipeline and data used in our analysis, and describe and justify the method followed when combining the different datasets, while we show and discuss our results in Sect.~\ref{sect:result}, and conclude in Sect.~\ref{sect:conclusions}. 
\section{Analysis and datasets}\label{sec:datamod}
Here we try to combine probes in the best that we can in order to obtain final constraints in a data driven model independent approach. We also want to avoid biases from the probes for which $H_0$ and $\sigma_8$ are showing tensions. In order to achieve that, we consider, either constraints obtained directly from measurements and not or weakly derived through a cosmological model, or probes that are made so if possible by relaxing their calibration or systematic nuisance parameters that are degenerate with $H_0$ or $\sigma_8$. As so, we will not include the direct local measurements on $H_0$ from Cepheid stars, nor measurements on $\Omega_{\rm{M}}$ and $\sigma_8$ from CMB angular power spectrum or BAO. The same does not apply for example for cluster counts for which we leave the mass observable calibration parameter as free relaxing by then their constraints on $\Omega_{\rm{M}}$ and $\sigma_8$ or to the luminosity distance from supernovae used later where we also let free the light curve calibration parameter. However, we can still adopt Gaussian prior on the power spectrum amplitude parameter $A_s$ and the spectral index $n_s$ from Planck 2018 (Plk18) CMB data \cite{Planck:2018vyg} and $\omega _{\text{b,}0}=0.0226\pm 0.00034$, as the average and standard deviation of a Gaussian prior on the baryon density, obtained by \cite{Cooke:2016rky} using big bang nucleosynthesis (BBN) + the abundance of primordial deuterium.\\ 

Then, as our geometric probe, we start by using $H(z)$ measurements that depends on the derivative of redshift with respect to cosmic time, known as cosmic chronometers (CC), following
\begin{equation}
H(z)= - \frac{1}{1+z} \frac{dz}{dt}.
\label{eq:Hubble}
\end{equation}
obtained by calculating the differential ages of passively evolving galaxies. This was introduced by \cite{moresco:2012by}, who proposed to use the break in the spectrum at 4000~\AA~rest-frame $D4000$, demonstrated to correlate extremely well with the stellar age and can be described by a simple linear relation:
\begin{equation}
D4000=A(Z, SFH)\times{\rm age}+B\; ,     
\label{eq:D4000age}
\end{equation}
where $B$ is a constant and  $A(Z, SFH)$ is a parameter, which depends only on the metallicity $Z$ and on the stellar function (SFH), and can be calibrated on stellar population synthesis models. By differentiating Eq.~\ref{eq:D4000age}, it is possible to derive the relation between the differential age evolution of the population and the differential evolution of the feature, in the form $dD4000=A(Z, SFH)\times{\rm dt}$ allowing us by then to decouple the statistical from the systematic effects, which results in the total covariance matrix for CC defined as:
\begin{equation}
{\rm Cov}_{ij}= {\rm Cov}_{ij}^{\rm stat}+ {\rm Cov}_{ij}^{\rm syst} \;\; ,
\end{equation}
where ${\rm Cov}_{ij}^{\rm syst}$, is decomposed into the several contributions mentioned above.

Here, we use the compilation of CC data points collected only by the above approach from Magana et al. \citep{Magana:2017usz} and Geng, et al. \citep{Geng:2018pxk} while removing other measurements of $H(z)$ obtained from e.g. BAO measurements even if we loose, by this procedure, some of the constraining power because we want to stay as model independent as possible. 

CC data will be part of our baseline but we also consider later a recent collection of measurements of luminosity distance from SNIa, known as the Pantheon sample~\cite{Pan-STARRS1:2017jku} distributed in the redshift interval $z \in [0.01, 2.3]$ where we leave its distance modulus calibration parameter $M_{\rm B}$ as free to vary.

While from the growth of structure sector side, we use the cluster counts probe relaxing the calibration parameter or other nuisances that could be degenerate with the latter. This is done in the present study using 
a Sunayev-Zeldovich (SZ) detected clusters sample \cite{2016A&A...594A..24P} where the distribution of clusters function of redshift and signal-to-noise is written as
\begin{eqnarray}
\frac{dN}{dz dq} =  \int d\Omega_{\rm mask} \int d{M_{500}} \, \left. \frac{dN}{dz d{M_{500}} d\Omega}\, \nonumber \right. \\ \left. \times P[q | {\bar{q}_{\rm m}}({M_{500}},z,l,b)] \right.
\end{eqnarray}
with
\begin{equation}
\label{eq:dndzdq}
\frac{dN}{dz d{M_{500}} d\Omega} = \frac{dN}{dV d{M_{500}}}\frac{dV}{dzd\Omega},
\end{equation}
and the quantity $P[q | {\bar{q}_{\rm m}}({M_{500}},z,l,b)]$ being the distribution of $q$ given the mean signal-to-noise value, ${\bar{q}_{\rm m}}({M_{500}},z,l,b)$, predicted by the model for a cluster of mass ${M_{500}}$ (i.e. defined at 500 the overdensity with respect to the critical density of the universe) and redshift $z$ located at the galactic coordinates $(l,b)$,
Which we relate to the measured integrated Compton $y$-profile $\bar{Y}_{500}$ using the following scaling relation :
\begin{equation}
\label{eq:Yscaling} 
E^{-\beta}(z)\left[\frac{{D_{\rm A}}^2(z) {\bar{Y}_{500}}}{\mathrm{10^{-4}\,Mpc^2}}\right] =  Y_\ast \left[ {\frac{h}{0.7}}
  \right]^{-2+\alpha} \left[\frac{(1-b)\,
    {M_{500}}}{6\times10^{14}\,M_{\odot}}\right]^{\alpha},
\end{equation}

where ${D_{\rm A}}$ is the angular diameter distance and
$E(z) = H(z)/H_0$, while  $\alpha$, $\beta$ and $Y_\ast$ are additional parameters in the SZ scaling law, along with $(1-b)$, that serves to link $M_{500}$ to $M_{\rm X}$, the cluster mass determined from X-ray observations, playing the role of the calibration parameter obtained from comparison with hydrodynamical simulations. Here we leave $(1-b)$ the calibration parameter and $\alpha$ that is weakly degenerate with the former as free to vary.\\
Another complementary probe to the cluster counts as well as to the geometric one introduced to help us break the degeneracies from previously relaxing the nuisance parameters is the gas mass fraction (GMF) probe which corresponds to 40 Chandra observations from massive and dynamically relaxed galaxy clusters in redshift range $0.078 \leq z \leq 1.063$ obtained by \cite{Mantz:2014xba}.
The gas mass fraction quantity for a cluster is given by \citep{Mantz:2014xba,Allen:2011zs}:
\begin{equation}
f_{\text{gas}}^{\text{X-ray}}\left( z\right) =A(z)K(z)\gamma (z) \frac{\Omega _{\text{b}}(z)}{\Omega _{\text{m}}(z)} \left[ \frac{D_{A}^{\text{fid}}\left( z\right) }{D_{A}\left( z\right) }\right] ^{\frac{3}{2}} \ , \label{Eq fgas}
\end{equation} 
where
\begin{equation}
    A(z)=\left[\frac{H(z)D_{A}(z)}{H^{\text{fid}}(z)D^{\text{fid}}_{A}(z)} \right]^{\eta}
\end{equation}
stands for the angular correction factor ($\eta=0.442\pm 0.035$), $\Omega_{\rm{m}}(z)$ is the total mass density parameter and  $\Omega_{\rm{b}}(z)$ the baryonic mass density parameter.  The parameters $\gamma(z)$ and $K(z)$  correspond, respectively, to the gaz depletion factor, and to the bias of X-ray hydrostatic masses due to both astrophysical and instrumental sources. 
By assuming $\omega _{\text{b,}0}\equiv \Omega _{\text{b,}0}h^{2}$, we can rewrite Equ.~(\ref{Eq fgas}) as
\begin{equation}
f_{\text{gas}}^{\text{X-ray}}\left( z\right) = 
\frac{K \ \gamma \ \omega _{\text{b,}0}}{\Omega _{\text{m,}0}h^2} \left[\frac{H(z)D_{A}(z)}{H^{\text{fid}}(z)D^{\text{fid}}_{A}(z)} \right]^{\eta}\left[ \frac{D_{A}^{\text{fid}}\left( z\right) }{D_{A}\left(z\right) }\right] ^{\frac{3}{2}}  .
\end{equation}
Therefore, for this sample, the $\chi ^{2}$ function is given by,
\begin{equation}
\chi_{\text{GMF}}^{2}=\sum_{i=1}^{40}\frac{\left[ f_{\text{gas}}^{\text{th}}(z_i)-f_{\text{gas}, \ i}^{\text{ob}}\right] ^{2}}{\sigma _{\text{tot}, \ i}^{2}}\text{ ,}
\end{equation}
with a total uncertainty given by 
\begin{eqnarray}
\sigma_{\text{tot}, i}^{2} = \sigma^{2}_{f_{\text{gas}}^{\text{ob}}, i} + \left[f_{\text{gas}}^{\text{th}}(z_i) \right]^{2} \left\{ \left( \frac{\sigma_{\Omega_b} }{\Omega_b}\right)^{2} + \left( \frac{\sigma_{\gamma }}{\gamma }\right)^{2} \nonumber \right. \\ \left.
   +  \ln^2 \left[\frac{H(z_i)D_{A}(z_i)}{H^{\rm{fid}}(z_i)D^{\rm{fid}}_{A}(z_i)}\right] \sigma_{\eta}^{2} \right\}
\end{eqnarray} 
We adopt the value of $\gamma=0.848 \pm 0.085$ in our analysis,\citep{Planelles:2014zaa,Mantz:2014xba}. The term in brackets corrects the angular diameter distance $D_A(z)$ from the fiducial model used in the observations, $D_A^\mathrm{fid}(z)$, which makes these measurements model-independent. It remains to relax the parameter $K(z)$ which we have left free since it is the one degenerate with the $\sigma_8$ tension.\\ 

Finally, to close our 'system' of constraints, we consider direct measurements of matter density by way of the Oort technique \cite{Oort:1958nna} that uses the mass to light $M/L$ ratio for galaxies in clusters divided by that of galaxies in the field $\rho_c/j$, where $j$ is the field luminosity density and $\rho_c$ the Universe critical density, as a direct measure of $\Omega_{\rm{M}}$ that is independent of the cosmology,
\begin{equation}
\Omega_{\rm{M}} = \frac{M/L}{\rho_c/j}.
\end{equation}

We compiled a list of the available observations to obtain direct measurements of $\Omega_{\rm{M}}$ in table \ref{tab:Om} from different studies with bounds that are compatible with each others. We note however that \cite{1998lsst.conf..119C} and \cite{Muzzin:2007rz} have worked on the same cluster sample, but they collected sources using a different waveband so that they capture a different population. Moreover, we checked that the results do not change significantly whether we combine both or choose either of them. We note also that there exist a measure of $M/L$ from Girardi \cite{Girardi:1999rp} that could be transformed, using the luminosity density of \cite{SDSS:2002vxn}, to also constraint $\Omega_{\rm{M}}$, but since their aim was not to provide a measure of the matter density, we did not add it but include it separately at the end of our table. However, here also we checked that the constraints inferred by the Monte Carlo Markov Chain MCMC runs with or without \cite{Girardi:1999rp} post processed data still yield the same results and conclusions we reach later in Sect.~\ref{sect:result}. \\

\begin{table}
\begin{tabular}{|lll|}
\hline
\quad $\Omega_{\rm{M}}$ & \quad $\sigma_{\Omega_{\rm{M}}}$ & \quad Reference  \\
\hline
 \quad 0.19 & \quad 0.06 & \quad Carlberg 1997 \cite{Carlberg:1995ra} \\
 \quad 0.16 & \quad 0.05 & \quad Bahcall 2000 \cite{Bahcall:2000zu} \\
 \quad 0.19 & \quad 0.03 & \quad Lin 2003  \cite{Lin:2003su} \\
 \quad 0.18 & \quad 0.03 & \quad Rines 2004 \cite{Rines:2004vg} \\
 \quad 0.22 & \quad 0.02 & \quad Muzzin 2007 \cite{Muzzin:2007rz}\\
 \quad 0.20 & \quad 0.03 & \quad Sheldon 2009 \cite{SDSS:2007umu}  \\
 \quad 0.26 & \quad 0.02 & \quad Bahcall 2014 \cite{Bahcall:2013epa} \\
 \hline
 \quad 0.20 & \quad 0.014 & \quad Girardi 2000 \cite{Girardi:1999rp} \\
\hline
\end{tabular}
\caption{Matter density measurements $\Omega_{\rm{M}}$ and their errors $\sigma_{\Omega_{\rm{M}}}$ and references of the works from where they were taken. The last point was used to test the robustness but was not included in the baseline MCMC analysis because originally only a mass to light ratio was provided and we used a luminosity density of \cite{SDSS:2002vxn} to determine $\Omega_{\rm{M}}$.}
\label{tab:Om}
\end{table}

We do not include local Hubble measurements nor weak lensing shear correlations measurements as previously mentioned because they are parts of the probes that are showing tensions and because it is not easy to find nuisance or calibration parameters that we could relax in the same way we followed for the above probes to make them less model dependent. Though BAO is also usually considered as the complement observation to CMB and both combinations agree on the high redshift values for $H_0$ and $\sigma_8$, hence it should not be included in our compilation of probes, however, being one of the strongest probe that forbid any correction from alleviating the tension we include it later only as an additional case to our baseline, the same as we shall do for the supernovae (SN) probe, to act as a robustness test when testing its impact on our results. Our BAO dataset will consist of 6DFGS \citep{Beutler:2011hx}, SDSS MGS \citep{Ross:2014qpa} and BOSS DR12 \citep{BOSS:2016wmc} and we refer to them in general as BAO.\\

We use MontePython, the cosmological Monte Carlo code \citep{Brinckmann:2018cvx} to estimate our parameters, in which we implemented or used the different described above likelihoods.

\begin{figure}[t]
	\centering
	\includegraphics[scale=0.5]{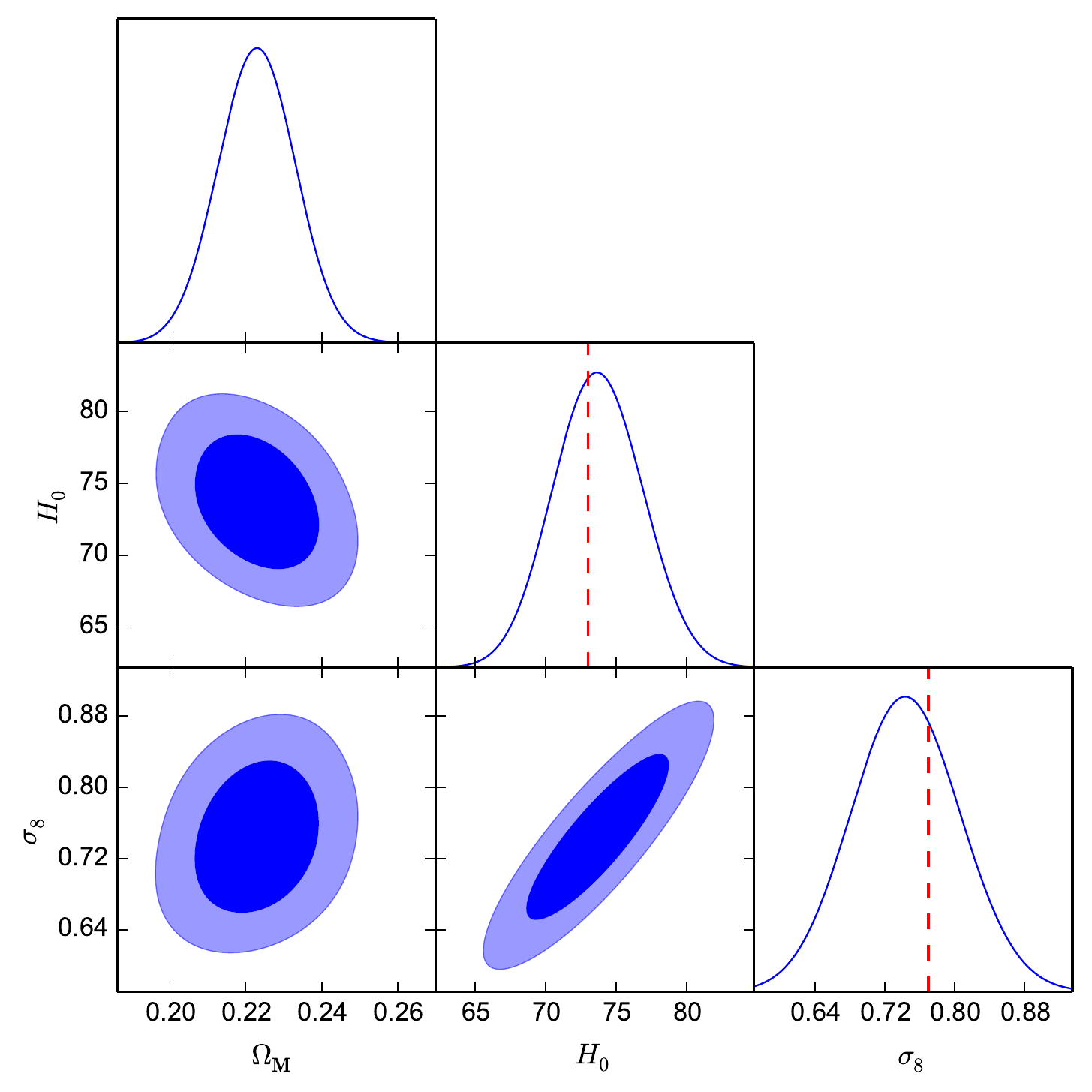}    
	\caption{ The 1D and 2D 68\% and 95\% confidence contours marginalised likelihood for the Hubble constant $H_0$ and $\sigma_8$, inferred from a combinations of cluster counts and gas fraction in galaxy clusters, cosmic chronometers and astrophysical constraints on $\Omega_{\rm{M}}$ with priors from BBN measurements as well as from CMB correlations on $n_s$ and $A_s$. The dashed lines corresponds to $H_0$ from local observations and $\sigma_8$ from weak lensing correlations and cluster counts when fixing their calibration using hydrodynamical simulations.}
	\label{fig:Om0chronoclustfrac}
\end{figure}
%
\section{Results and discussion}\label{sect:result}

\begin{figure}[h]
	\centering
	\includegraphics[width=\columnwidth]{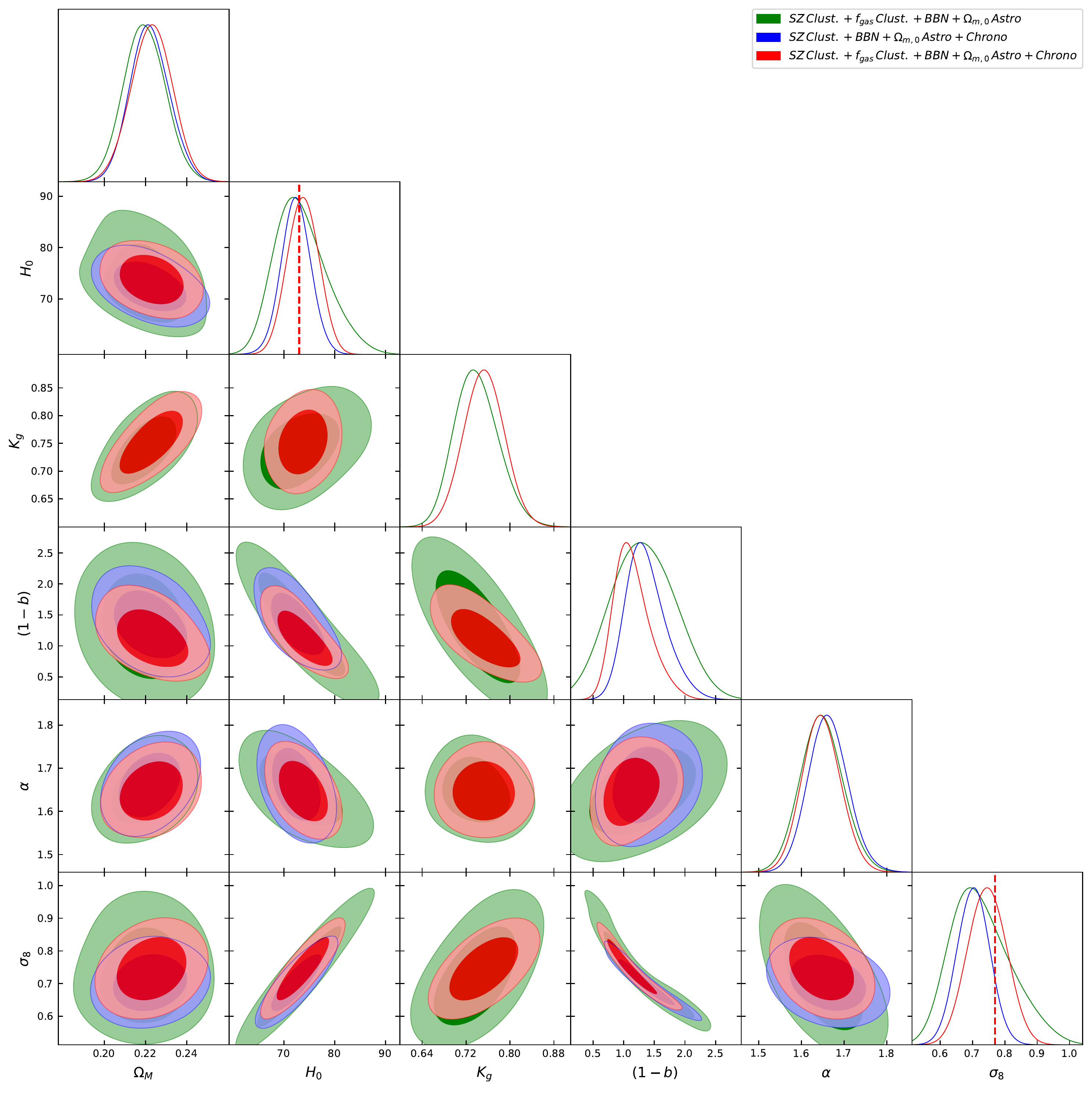}  	
	\caption{The 1D and 2D 68\% and 95\% confidence contours marginalised likelihood for the Hubble constant $H_0$ and $\sigma_8$ , inferred using the same probes as in Fig.~\ref{fig:Om0chronoclustfrac} but each time taking out one probe to highlight the necessity of combining all and the complementarity between them (see legend for details) . The dashed lines corresponds to $H_0$ from local observations and $\sigma_8$ from weak lensing correlations and cluster counts when fixing their calibration using hydrodynamical simulations.}
	\label{fig:suball}
\end{figure}

\begin{figure}
	\centering
          \includegraphics[scale=0.4]{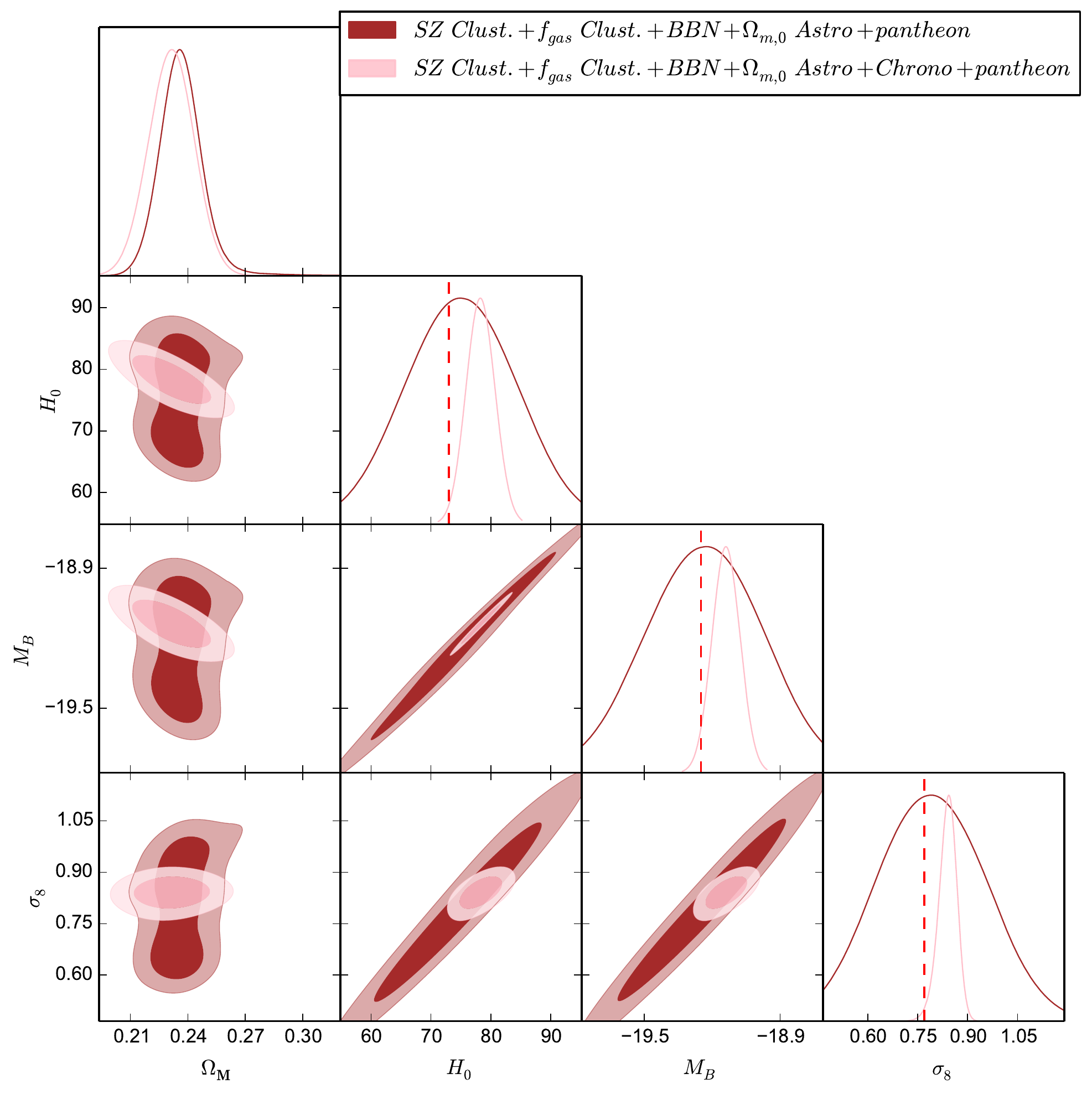}
          
	\caption{The 1D and 2D 68\% and 95\% confidence contours marginalized likelihood for the Hubble constant $H_0$ and $\sigma_8$, inferred using cluster counts, gas fraction in galaxy clusters, Pantheon supernova sample and astrophysical constraints on $\Omega_{\rm{M}}$ with priors from BBN measurements (red lines) and for the case where the cosmic chronometers were added (pink lines).}
	\label{fig:PanthCC}
\end{figure}

\begin{figure}
	\centering
    \includegraphics[scale=0.4]{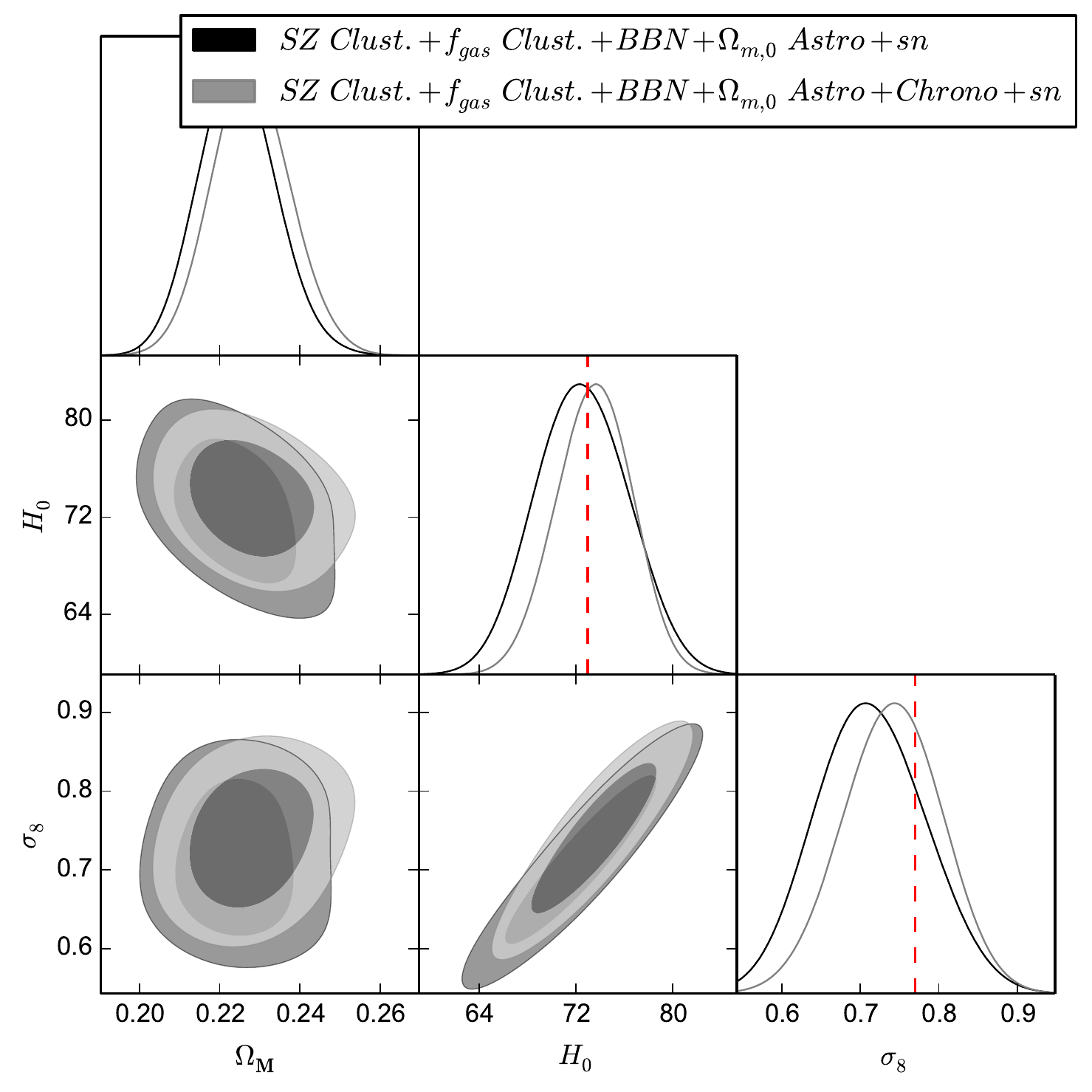}
	\caption{The 1D and 2D 68\% and 95\% confidence contours marginalized likelihood for the Hubble constant $H_0$ and $\sigma_8$, inferred using cluster counts, gas fraction in galaxy clusters, supernova sample from \cite{2010ApJ...716..712A} and astrophysical constraints on $\Omega_{\rm{M}}$ with priors from BBN measurements (black lines) and for the case where the cosmic chronometers were added (gray lines).}
	\label{fig:SNCC}
\end{figure}

\begin{figure}
	\centering
    \includegraphics[scale=0.35]{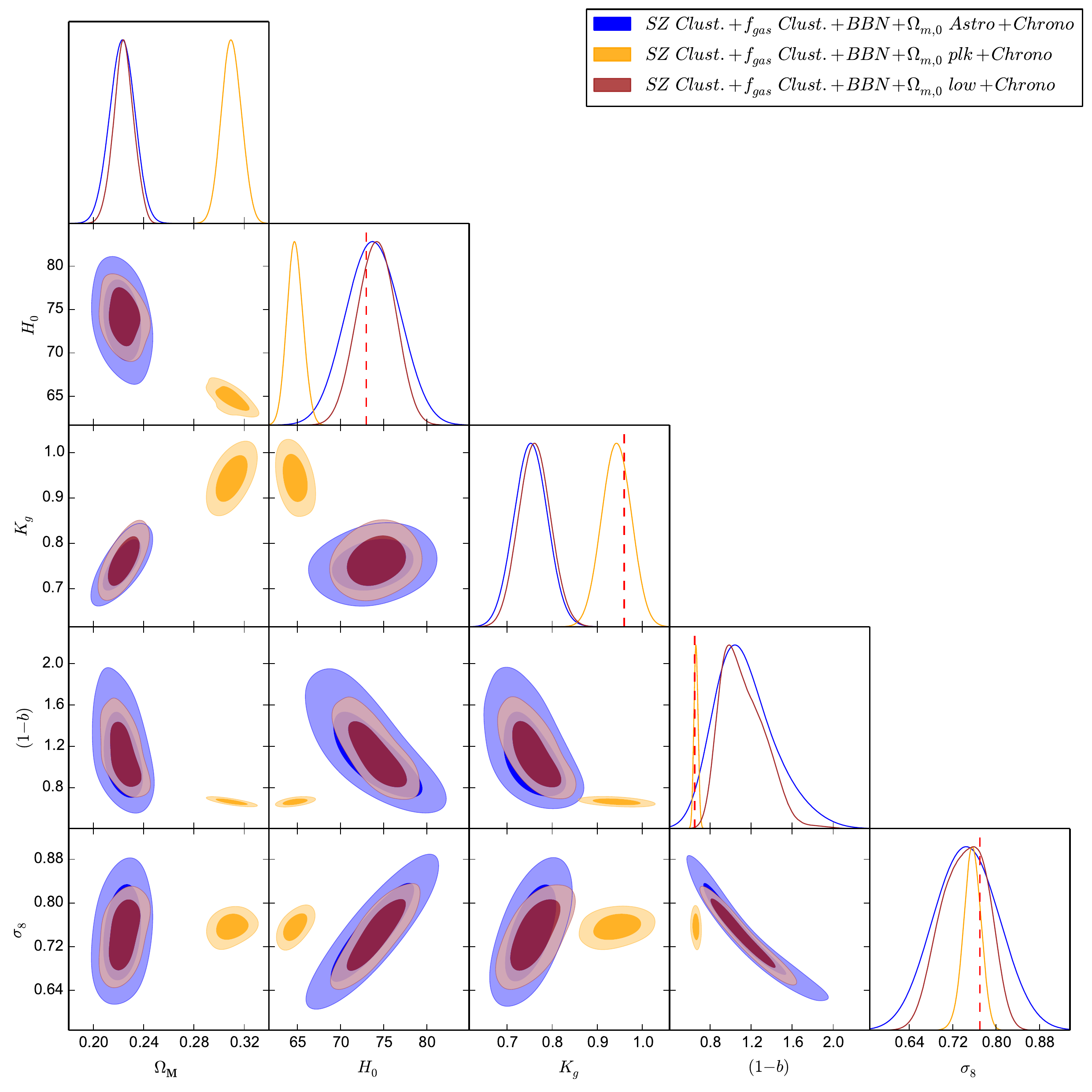}
	\caption{The 1D and 2D 68\% and 95\% confidence contours marginalized likelihood for the Hubble constant $H_0$ and $\sigma_8$, inferred using cluster counts, gas fraction in galaxy clusters, cosmic chronometers and astrophysical constraints on $\Omega_{\rm{M}}$ with priors from BBN measurements (blue lines) and for the case where the matter density measurements were replaced by a prior from Plk18 (yellow lines) or by a Gaussian prior on the matter density with average the best value from the astrophysical constraints while the standard deviation matches the one from Plk18 (red lines).}
	\label{fig:Om0chronoOmplk}
\end{figure}

We start by showing in Fig.~\ref{fig:Om0chronoclustfrac} constraints on the three parameters subject of discrepancy $H_0$, $\Sigma_8$ and $\Omega_{\rm{M}}$, inferred from MCMC runs following the method detailed on in the previous section, using first, considered as our baseline, a combination of the cosmic chronometers, the gas fraction in galaxy clusters, the clusters counts, the big bang nucleosynthesis and the direct measurements of the matter density. We leave free the $\Omega_{\rm M}$, $h$ and $\sigma_8$ cosmological parameters, and the relevant degenerate nuisance parameters as well, namely $(1-b)$ and $K_g$ for the cluster based probes. We also adopt priors from CMB and BBN data on the remaining cosmological parameters, $n_s$, $A_s$ and $\Omega_b$. We observe that this combination yields constraints for the Hubble constant $H_0$ and $\sigma_8$ compatible with those usually found from local probes while that on the value of the matter density is more than 3$\sigma$ in discrepancy with the one usually obtained from Plk18 data, suggesting that a low value for the matter density is compatible with both discrepant parameters, $H_0$ and $\sigma_8$ local inferred values.
To better understand the contribution and role of each probe and the need to include it to break degeneracies in our model independent-like approach, we show in the following how different sub-combinations of the datasets used would constrain the evolution of our free parameters. Thus in Fig.~\ref{fig:suball} we compare our baseline with the case where we omit, either the cluster gas fraction, or the SZ cluster counts or the CC probe. We observe that by omitting only the cosmic chronometers probe (green lines) or cluster gas fraction (blue lines), we are still in agreement with our findings, however the contours of $H_0$ and $\sigma_8$ widen for the former case while $\sigma_8$ contours are shifted to low values in the latter becoming even not compatible with the local weak lensing constraints form \cite{DES:2021wwk} or \cite{KiDS:2020ghu} while the galaxy clusters bias is driven higher but to values way above those usually obtained from clusters when calibrated by hydrodynamical simulations \cite[see][for example and reference therein]{2014A&A...571A..20P}.
Next we shall continue to test the robustness of our findings by further replacing some of our datasets by those from supernovae (SN) luminosity distance probe in order to observe the impact of such change on our previous bounds.
%
As so we show in Fig.~\ref{fig:PanthCC} the results from further adding the supernovae sample Pantheon \cite{Pan-STARRS1:2017jku}, since one could argue that this sample contains sources that spans a large range from low $z \sim 0$ to high redshifts $z\sim 2.5$ while being at the same time free from the BAO+CMB constraining connection. To enter our criteria and further free the probe from the Cepheids' calibration prior, we let its calibration $M_B$ free to vary.
However we see that its constraining power is strongly decreased when a free $M_B$ (red lines) is adopted and our combination with the CC is not included, rendering the constraints on $H_0$ and $\sigma_8$ very loose, though we observe that the maximum likelihood still prefers values in agreement with their local bounds. While when keeping the cosmic chronometers along with the SN datasets, thus further breaking degeneracies, we observe that the Pantheon sample, is pushed by the low bounds of $\Omega_{\rm{M}}$ previously found, to choose instead higher values for $H_0$ and $\sigma_8$. This is probably because the augmented version Pantheon+ was shown in \cite{Brout:2022vxf} to prefer values of $\Omega_{\rm{M}}$ slightly higher than those from CMB datasets thus the counter adjustment seen here. We note that this could also signify that Pantheon or Pantheon+ is in discrepancy with other probes and suffer from possible misdetermination of systematics as noted in some studies \cite{2022arXiv221207917K,Perivolaropoulos:2023iqj,Dainotti:2021pqg,2022arXiv220611447C}. As a test also on that possibility, we have rerun MCMC using older supernova data from \cite{2010ApJ...716..712A}. We observe in Fig.~\ref{fig:SNCC} a very good agreement and compatibility with our baseline although for this SN dataset the calibration parameter was fixed to that obtained from Cepheid stars and was not left free to vary. Nevertheless we note that an investigation using Pantheon+ should be performed with the same combinations as here, but we leave this test to future studies, since the MCMC runs of this work were already in an advanced stage when the Pantheon+ was released. 
%
Finally, we finish our runs with one where we use the same baseline combination but replacing the constraint on $\Omega_{\rm{M}}$ by a prior from \cite{Planck:2018vyg}. We observe, as seen in Fig.~\ref{fig:Om0chronoOmplk} with the yellow lines that an $H_0$ close to the usual value constrained by CMB+BAO is preferred again while $\sigma_8$ remains within the constraints we obtain from weak lensing or cluster counts probe. However, since $\sigma_8$ here seems to be fixed by the two used galaxy clusters probes, the compensation of choosing a high value for $\Omega_{\rm{M}}$ translates in a shift in the value of their calibration parameters $(1-b)$ and $K_g$ to values needed to alleviate the tension between CMB and clusters, in agreement with what was found and noted by \cite{2016A&A...594A..24P} and \cite{Sakr:2018new}. This is further confirmed if we consider now a prior on $\Omega_{\rm{M}}$ with a small value around the maximum likelihood previously obtained from the direct matter density sample but now with bounds as strong as the ones we usually obtain from CMB. We see (pink lines) that $H_0$ matches that of \citep{2022ApJ...934L...7R} while the calibration parameters of the cluster probes shift back to the values found by the SZ Planck collaboration when they are calibrated based on hydrodynamical simulations confirming how future better direct measurements of the matter density could be used to confirm or rule out our proposed $\Omega_{\rm M}$ tension.

%
\section{Conclusion}\label{sect:conclusions}
In this work we wanted to test the hypothesis on whether the discrepancies on $H_0$ and $\sigma_8$ are compatible with the existence of an additional one on the matter density. For that we performed a Bayesian analysis on the cosmological parameters using and combining several probes in an agnostic way by, either relaxing some of their nuisance parameters that could be degenerate with the parameters subject of tension such as the mass observable calibration parameter for cluster counts, or by only considering Gaussian priors on parameters they directly measure such as the spectral index from CMB power spectrum, or simply because they are weakly dependent of the parameters subject of discrepancy such as the cosmic chronometers, all in the final aim to break degeneracies and auto calibrate the free non informative nuisance parameters as well as the ones subject of discrepancies. Since usually $\Omega_{\rm{M}}$ is strongly determined from CMB with or without BAO while we wanted to test how sensitive it is to $H_0$ and $\sigma_8$ outside the constraints from CMB, we used a sample of direct measures of $\Omega_{\rm{M}}$ obtained by comparing mass to light ratio in clusters over that from galaxies in the background field.\\
We found that our combination yields a low matter density value, as expected since the measurements of $\Omega_{\rm{M}}$ we used are all much below the value inferred by CMB, but also an $H_0$ and $\sigma_8$ compatible with values obtained from either local or cosmological free probes, i.e. weak lensing or measurements of the Hubble parameter from Cepheid stars. We also had similar results albeit with a widening of the constraints when we replace the cosmic chronometer sample by the supernovae recent compiled sample from Pantheon while leaving its calibration parameter free. However, combining with both chronometers and Pantheon shifted $H_0$ and $\sigma_8$ to higher values due to the fact that Pantheon prefers usually even higher values than CMB for $\Omega_{\rm{M}}$ while our sample put constrain on this parameter which ultimately translates into these shifts in $H_0$ and $\sigma_8$. We note however that this is observed with recent SN measurements while using older SN data resulted in no change in our baseline combination whether alone or combined with cosmic chronometers.\\
\begin{figure}
	\centering
	\includegraphics[scale=0.5]{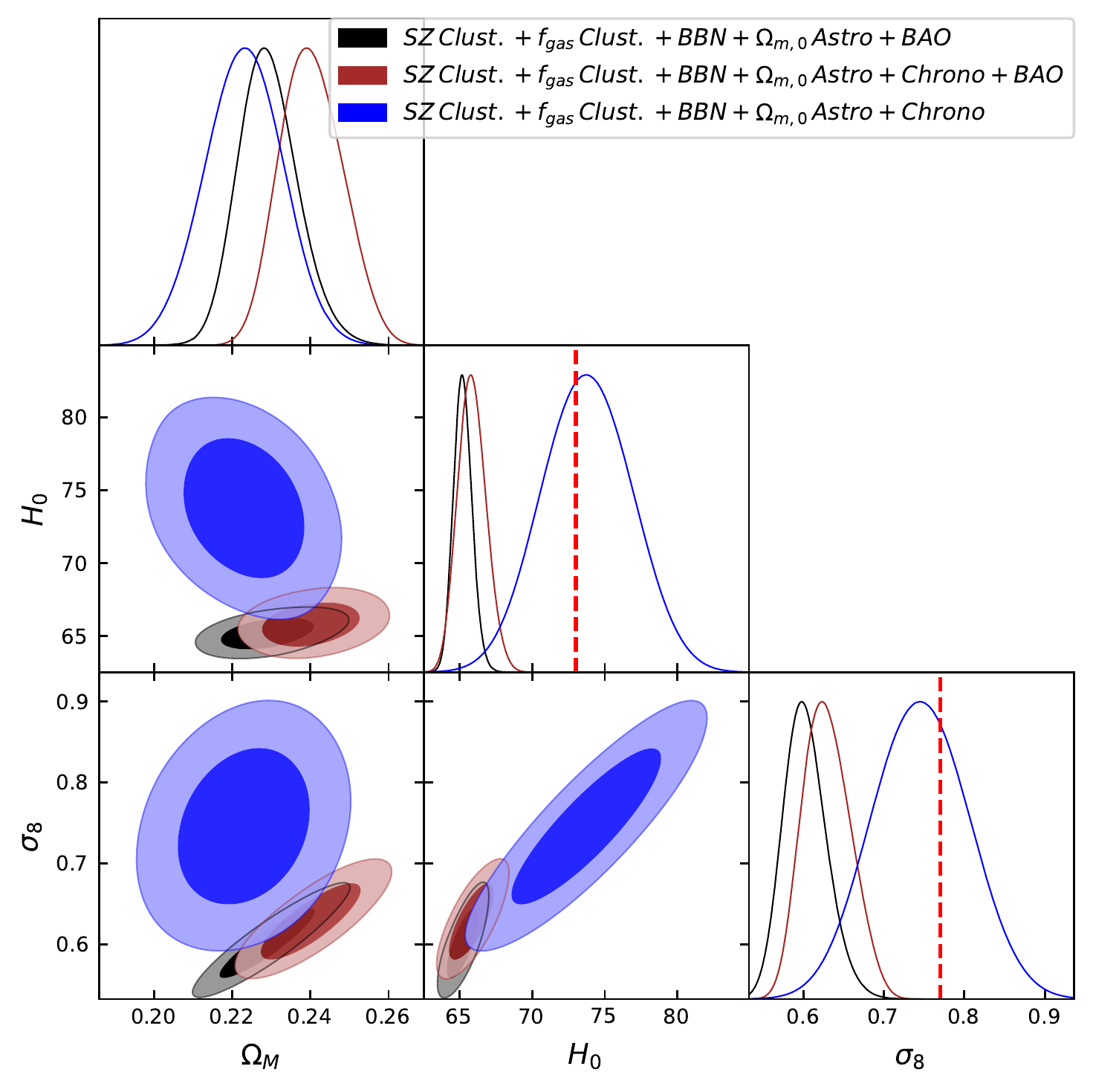} 
	\caption{The 1D and 2D 68\% and 95\% confidence contours marginalized likelihood for the Hubble constant $H_0$ and $\sigma_8$, inferred using cluster counts, gas fraction in galaxy clusters, BAO and astrophysical constraints on $\Omega_{\rm{M}}$ with priors from BBN measurements (black lines) and for the case where the cosmic chronometers were added (red lines) vs the basline without BAO probe (blue lines).}
	\label{fig:Om0ChronoBAO}
\end{figure}
We ended by a test in which we replaced the local $\Omega_{\rm{M}}$ constraints by a prior from CMB Planck \cite{Planck:2018vyg} or by a prior with low values of $\Omega_{\rm{M}}$ but with tighter constraint similar to those obtained from CMB. We found that the combination with BAO without cosmic chronometers or the change in the value of $\Omega_{\rm{M}}$ translates in that of $H_0$ from values compatible with CMB to those in agreement with local $H_0$ while we observe at the same time a shift in the calibration parameters, usually degenerate with $\sigma_8$, from values that are compatible with CMB to those in agreement with clusters calibration obtained from hydrodynamical simulations.\\
And as a further stress we run and show in Fig.~\ref{fig:Om0ChronoBAO} MCMC results when adding BAO, despite that this probe does not match our considered criteria for including it in our collection of probes since its $H_0$ inferred value is showing the same discrepancy as is the case for the CMB vs local ones, and it is difficult to find and relax a calibration parameter that might be responsible for the difference. We observe that $H_0$ is restored back to $\sim 67.0$ in the case when we combine with our baseline probes and that while keeping or omitting CC. This is due to the fact that BAO still have its full constraining power from being used in a model dependent way. However, we observe that $\sigma_8$ is severely pushed to low values showing the non compatibility of high redshift or CMB compatible probes with local measures, including a local $\Omega_{\rm{M}}$, while when using CC instead, the constraints on $\sigma_8$ are shifted a little below its local values.\\
We conclude that local measurements yielding a discrepancy on the matter density could be  added to the list of existing tensions within $\Lambda$CDM model. At best our results are pointing to a problem between different probes and the way measurements are performed or the assumptions used when extracting data, but also this could be indicating that $\Lambda$CDM model is facing further troubles in accommodating all the existing probes at once. \\

\begin{acknowledgments}
We thank the referee for their valuable suggestions and improvements to the manuscript. Z.S. acknowledges funding from DFG project No. 456622116 and support from the IRAP Toulouse and IN2P3 Lyon computing centres.
\end{acknowledgments}



\bibliography{apssamp}

\end{document}